\documentclass[aps,floatfix,twocolumn]{revtex4}

\usepackage{amsmath}
\usepackage{amsfonts}
\usepackage{amssymb}
\usepackage{graphics}

\begin{document}

\title{Decoherence-induced geometric phase in a multilevel atomic
system}
\author{Shubhrangshu Dasgupta$^1$ and Daniel A. Lidar$^{1,2}$}
\affiliation{$^1$Department of Chemistry, University of Southern California, Los Angeles,
CA 90089, USA\\
$^{2}$Departments of Electrical Engineering and Physics, University of
Southern California, Los Angeles, CA 90089, USA}
\date{\today}

\begin{abstract}
We consider the STIRAP process in a three-level atom. Viewed as a closed
system, no geometric phase is acquired. But in the presence of spontaneous
emission and/or collisional relaxation we show numerically that a
non-vanishing, purely real, geometric phase is acquired during STIRAP, whose
magnitude grows with the decay rates. Rather than viewing this
decoherence-induced geometric phase as a nuisance, it can be considered an
example of \textquotedblleft beneficial decoherence\textquotedblright :\ the
environment provides a mechanism for the generation of geometric phases
which would otherwise require an extra experimental control knob.
\end{abstract}

\maketitle

\section{Introduction}

Berry observed that quantum systems may retain a memory of their
motion in Hilbert space through the acquisition of geometric
phases~\cite{Berry:84}. Remarkably, these phase factors depend only on
the geometry of 
the path traversed by the system during its evolution. Soon after this
discovery, geometric phases became a subject of intense theoretical and
experimental studies~\cite{Wilczek:book}. In recent years, renewed interest
has arisen in the study of geometric phases in connection with quantum
information processing~\cite{Zanardi:99b,Jones:00}. Indeed, geometric, or
holonomic quantum computation (QC) may be useful in achieving fault
tolerance, since the geometric character of the phase provides protection
against certain classes of errors~\cite%
{Solinas:04,Zhu:05,Fuentes-Guridi:05,WuZanardiLidar:05}. However, a
comprehensive investigation in this direction requires a generalization of
the concept of geometric phases to the domain of \emph{open} quantum
systems, i.e., quantum systems which may decohere due to their interaction
with an external environment.

Here we consider the following basic question:

\begin{quote}
Is it possible for the environment to induce a geometric phase where there
is none if the system is treated as closed?
\end{quote}

Apart from its fundamental nature, this question is of obvious practical
importance to holonomic QC, since if the anwer is affirmative the
corresponding open-system geometric phase can either be detrimental (if it
causes a deviation from the intended value) or beneficial, in the sense that
the environment is acting as an amplifier for, or even generator of, the
geometric phase.

Geometric phases in open systems, and more recently their applications in
holonomic QC, have been considered in a number of works, since the late
1980's. The first, phenomenological approach to the subject used the Schr%
\"{o}dinger equation with non-Hermitian Hamiltonians~\cite%
{Garrison:88,Dattoli:90}. While a consistent non-Hermitian Hamiltonian
description of an open system in general requires the theory of stochastic
Schr\"{o}dinger equations \cite{Gardiner:book}, this phenomenological
approach for the first time indicated that complex Abelian geometric phases
should appear for systems undergoing cyclic evolution. In Refs.~\cite%
{Ellinas:89,Gamliel:89,Romero:02,Whitney:03,Whitney:05,Kamleitner:04},
geometric phases acquired by the density operator were analyzed for various
explicit models within a master equation approach. In Refs.~\cite%
{Carollo:03,Fuentes-Guridi:05}, the quantum jumps method was employed to
provide a definition of geometric phases in Markovian open systems (related
difficulties with stochastic unravellings have been pointed out in Ref.~\cite%
{Bassi:05}). In another approach the density operator, expressed in its
eigenbasis, was lifted to a purified state \cite{Tong:04,Rezakhani:05}. In
Ref.~\cite{Marzlin:04}, a formalism in terms of mean values of distributions
was presented. An interferometric approach for evaluating geometric phases
for mixed states evolving unitarily was introduced in Ref.~\cite{Vedral:00}
and extended to non-unitary evolution in Refs.~\cite{Faria:03,Ericsson:03}.
This interferometric approach can also be considered from a purification
point of view~\cite{Vedral:00,Ericsson:03}. This multitude of different
proposals revealed various interesting facets of the problem. Nevertheless,
the concept of adiabatic geometric phases in open systems remained
unresolved in general, since most of these treatments did not employ an
adiabatic approximation genuinely developed for open systems. Note that the
applicability of the closed systems adiabatic approximation~\cite%
{Messiah:vol2} to open systems problems is not a priori clear and should be
justified on a case-by-case basis. Moreover, almost all of the previous
works on open systems geometric phases were concerned with the Abelian
(Berry phase) case. Exceptions are the very recent Refs.~\cite%
{Fuentes-Guridi:05,Florio:06,Trullo:06}, which discuss both non-adiabatic
and adiabatic dynamics, but employ the standard adiabatic theorem for closed
systems in the latter case.

Recently, a fully self-consistent approach for both Abelian and non-Abelian
adiabatic geometric phases in open systems was proposed by Sarandy and Lidar
(SL) in Ref.~\cite{SarandyLidar:06}. It applies to the very general class of
systems described by convolutionless master equations \cite{Breuer:book}. SL
made use of the formalism they developed in Ref.~\cite{SarandyLidar:04} for
adiabaticity in open systems, which relies on the Jordan normal form of the
relevant Liouville (or Lindblad) super-operator. The geometric phase was
then defined in terms of the left and right eigenvectors of this
super-operator. This definition is a natural generalization of the one given
by Berry for a closed system, and was shown to have a proper closed system
limit. The formalism was illustrated in Ref.~\cite{SarandyLidar:06} in the
context of a spin system interacting with an adiabatically varying magnetic
field.

In order to address the basic question posed above, we study here the
adiabatic geometric phase in a multi-level atomic system using the SL
formalism. Specifically, we consider the process of stimulated Raman
adiabatic passage (STIRAP) \cite{Bergmann:98,Unanyan:99} in a three-level
atomic system in a $\Lambda $ configuration. We analyze a version of STIRAP
where the closed system geometric phase is identically zero. We then show
that when spontaneous emission and/or collisional relaxation are included,
the same STIRAP process yields a non-vanishing geometric phase. This
decoherence-induced geometric phase is an example of 
``beneficial decoherence'', where the environment performs a
potentially useful task. This is conceptually similar to
the phenomenon of decoherence-induced entanglement
\cite{Plenio:99,PhysRevA.65.040101}.

Since the SL
formalism involves finding the Jordan normal form of a general matrix, which
is an analytically difficult problem, we developed a numerically stable
program to find the Jordan form of any complex square matrix and used it to
find the geometric phase \cite{program}.

The structure of the paper is as follows. In Section~\ref{sec2} we briefly
review the STIRAP process in a closed three level system in the $\Lambda $
configuration, and the corresponding calculation of the (vanishing)
geometric phase. In Section~\ref{sec3} we revisit this problem in the open
system setting and derive the solution of the STIRAP model. Our numerical
results, along with a detailed analysis of the geometric phase, are
presented in Section~\ref{results}. We conclude in Section~\ref{conc}.

\section{Geometric Phase under STIRAP:\ The Closed System Case}

\label{sec2}

\begin{figure}[tbp]
\scalebox{0.4}{\includegraphics{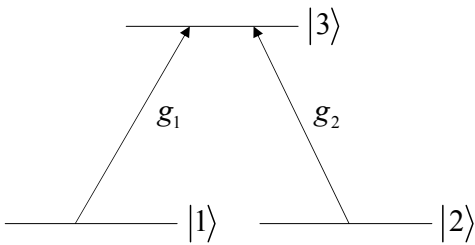}}  \vspace{-4.5cm}
\caption{Three-level atomic configuration with degenerate ground state
levels $|1\rangle $, $|2\rangle $, and excited state $|3\rangle $. The atom
interacts with two resonant classical fields with time-dependent Rabi
frequencies $g_{1}(t)$ (probe laser) and $g_{2}(t)$ (Stokes laser). }
\label{fig:3LS}
\end{figure}

We consider the process of stimulated Raman adiabatic passage (STIRAP) \cite%
{Bergmann:98} in a three-level system in the $\Lambda $ configuration, as
shown in Fig.~\ref{fig:3LS}. 
In this process, the initial atomic population in level $|1\rangle $ is
completely transferred to level $|2\rangle $, while the pulses are applied
in a \textquotedblleft counterintuitive\textquotedblright\ sequence. The
intermediate level $|3\rangle $ does not become substantially populated. The
interaction picture Hamiltonian in one-photon resonance can be written in
the rotating-wave approximation as follows: 
\begin{equation}
H=g_{1}(t)|3\rangle \langle 1|+g_{2}(t)|3\rangle \langle 2|+\mathrm{h.c.},
\label{hamil}
\end{equation}%
where the real functions $g_{i}$ are the time-dependent Rabi frequencies of
the two laser pulses, interacting respectively with the transitions $%
|i\rangle \leftrightarrow |3\rangle $ ($i\in 1,2$). The eigenvalues of $H$
are given by 
\begin{equation}
E_{0}=0,E_{\pm }=\pm \sqrt{g_{1}^{2}+g_{2}^{2}}  \label{eq:E}
\end{equation}%
and the respective eigenvectors are given by 
\begin{eqnarray}
|0\rangle &=&\cos (\theta )|1\rangle -\sin (\theta )|2\rangle ,  \notag \\
|+\rangle &=&\sin (\theta )\sin (\phi )|1\rangle +\cos (\theta )\sin (\phi
)|2\rangle +\cos (\phi )|3\rangle ,  \notag \\
|-\rangle &=&\sin (\theta )\cos (\phi )|1\rangle +\cos (\theta )\cos (\phi
)|2\rangle -\sin (\phi )|3\rangle ,  \notag \\
&&  \label{adia}
\end{eqnarray}%
where $\tan (\theta )=g_{1}/g_{2}$. Thus the time-dependence of the
eigenfunctions is parameterized by that of $\theta $. In principle the $%
g_{i} $'s can be complex valued, which gives rise to a controllable phase $%
\phi $ \cite{WuZanardiLidar:05}. Here we work with real valued $g_{i}$'s and
set $\phi =\pi /4$ for the remainder of this work. The state $|0\rangle $ is
a dark state, i.e., it has eigenvalue $0$.

We choose a Gaussian time-dependent profile for the control pulses: 
\begin{equation}
g_{1}(t)=g_{01}e^{-(t-t_{0})^{2}/\tau ^{2}},\;g_{2}(t)=g_{02}e^{-t^{2}/\tau
^{2}},  \label{pulse}
\end{equation}%
where $g_{01}$ and $g_{02}$ are the pulse amplitudes, and $t_{0}$ is the
time-delay between the pulses, with pulse $g_{2}$ preceding pulse $g_{1}$.
All time-scales are normalized in terms of the pulse-width $\tau $. The
closed-system adiabaticity condition is satisfied provided $t_{0}\sim \tau $
and \cite{Bergmann:98} 
\begin{equation}
\frac{\left\vert \frac{\partial \theta }{\partial t}\right\vert }{\sqrt{%
g_{1}(t)^{2}+g_{2}(t)^{2}}}\ll 1\quad \forall t.  \label{eq:adcond}
\end{equation}%
In this limit, the evolution of the system strictly follows the evolution of
either of the adiabatic states. Due to the ordering of the pulses as in (\ref%
{pulse}), the atom initially in the level $|1\rangle $ is prepared in the
adiabatic state $|0\rangle $. The population in level $|1\rangle $ is then
completely transferred to level $|2\rangle $ adiabatically, following the
evolution of state $|0\rangle $ under the action of the pulses (\ref{pulse}%
). Note that as the system follows the evolution of the state $|0\rangle $
in\ the adiabatic limit and the excited level $|3\rangle $ does not
contribute to $|0\rangle $, the traditional view of the process is that it
remains unaffected by spontaneous emission. Below we will show how this view
must be modified in a consistent treatment of the process as evolution of an
open system. In addition, incoherent processes such as dephasing of the
ground state levels will affect the population transfer process.

\begin{figure}[tbp]
\scalebox{0.3}{\includegraphics{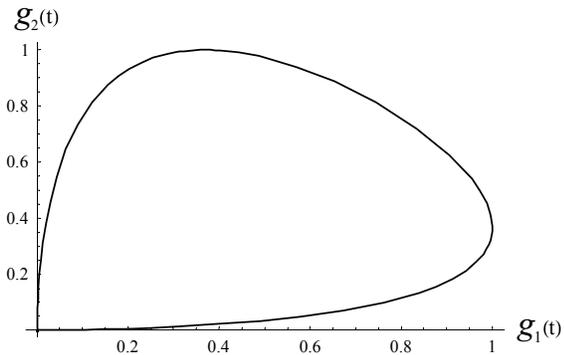}}
\caption{A closed curve in the $(g_{1},g_{2})$ parameter space, for $t_{0}=%
\protect\tau =g_{01}=g_{02}=1$. At $t=-\infty $ the curve is at the origin,
then rises steeply, and eventually returns to the origin at $t=+\infty $.}
\label{fig:g1g2}
\end{figure}

The geometric phases acquired by each adiabatic state $|n\rangle $, as
acquired during the evolution between $t_{0}$ and $t$ can be easily
calculated from \cite{Berry:84} 
\begin{equation}
\beta _{n}=i\int_{t_{0}}^{t}dt^{\prime }\langle n|\frac{d}{dt^{\prime }}%
|n\rangle .
\end{equation}%
In terms of a vector $\vec{R}(t)$ in parameter space undergoing cyclic
evolution, this phase can be rewritten as 
\begin{equation}
\beta _{n}=i\oint \langle n(\vec{R})|\frac{d}{d\vec{R}}|n(\vec{R})\rangle
\cdot d\vec{R}.  \label{eq:Berry}
\end{equation}%
In the case of the three-level system depicted in Fig.~\ref{fig:3LS}, the
parameter space is defined by $g_{1}(t)$ and $g_{2}(t)$, i.e., 
\begin{equation}
\beta _{n}=i\sum_{j=1}^{2}\oint \langle n(g_{1},g_{2})|\frac{\partial }{%
\partial g_{j}}|n(g_{1},g_{2})\rangle dg_{j}.  \label{eq:g1g2}
\end{equation}
We consider a cyclic evolution in this parameter space, which takes place as 
$t$ varies from $-\infty $ to $+\infty $, i.e., $g_{1}(-\infty
)=g_{2}(-\infty )=0$, $g_{1}(\infty )=g_{2}(\infty )=0$. This is shown in
Fig.~\ref{fig:g1g2}. 
One can also parametrize the time-dependence of the pulses in terms of an
angle $\theta $, using (\ref{pulse}), such that 
\begin{equation}
\tan \theta (t)=\frac{g_{1}(t)}{g_{2}(t)}=\frac{g_{01}}{g_{02}}%
e^{(2tt_{0}-t_{0}^{2})/\tau ^{2}}.
\end{equation}%
Then as $t$ varies from $-\infty $ to $+\infty $ we have that $\tan \theta
(t)$ varies from $0$ to $\infty $, and hence $\theta (t)$ varies from $0$ to 
$\pi /2$. Changing variables in Eq.~(\ref{eq:Berry}), the geometric phase
becomes in our case:%
\begin{equation}
\beta _{n}=i\int_{0}^{\pi /2}\langle n(\theta )| \frac{d}{d\theta } |
n(\theta )\rangle d\theta .
\label{eq:theta}
\end{equation}%
Note that the relevant parameter space for our problem is that with
coordinates $(g_{1},g_{2})$, not $(\theta ,\varphi )$ of Eq.~(\ref{adia});
indeed, Eq.~(\ref{eq:theta}) does not even describe a cycle in the $(\theta
,\varphi )$ space, whereas the expression~(\ref{eq:g1g2}) along with Fig.~%
\ref{fig:g1g2} show clearly that there is a cycle in the $(g_{1},g_{2})$
space.

Let us now show that the geometric phase vanishes for all three adiabatic
eigenstates $|n\rangle $ of (\ref{adia}), because the integrand $\langle
n(\theta )|\frac{d}{d\theta }|n(\theta )\rangle \equiv 0$. Indeed, consider
the adiabatic eigenstates of Eq.~(\ref{adia}). Then $\langle +|\frac{d}{%
d\theta }|+\rangle =\frac{1}{2}(\sin (\theta )\langle 1|+\cos (\theta
)\langle 2|+\cos (\phi )\langle 3|)(\cos (\theta )|1\rangle -\sin (\theta
)|2\rangle ) 
=0$, and $\langle -|\frac{d}{d\theta }|-\rangle =\frac{1}{2}(\sin (\theta
)\langle 1|+\cos (\theta )\langle 2|-\sin (\phi )\langle 3|)(\cos (\theta
)|1\rangle -\sin (\theta )|2\rangle ) 
=0$, irrespective of the value of $\phi $. Also, $\langle 0|\frac{d}{d\theta }%
|0\rangle =(\cos (\theta )\langle 1|-\sin (\theta )\langle 2|)(-\sin (\theta
)|1\rangle -\cos (\theta )|2\rangle )=0$. Thus the STIRAP process under
consideration does not give rise to a closed-system geometric phase. We note
that the analysis above is a special case of the four-level model considered
in Ref.~\cite{Unanyan:99}.

\section{Geometric Phase under STIRAP:\ The Open System Case}

\label{sec3}

\subsection{The model}

We now analyze the effect on the geometric phase of interaction of the
atomic system with a bath causing spontaneous emission and collisional
relaxation. We describe these processes in the Markovian limit for the bath,
using time-independent Lindblad operators and neglecting Lamb and Stark
shift contributions \cite{Breuer:book}. Thus the time-dependence appears
only in the control Hamiltonian $H$ [Eq.~(\ref{hamil})], and the evolution
of the system density matrix $\rho $ is given by the Lindblad equation (in $%
\hbar =1$ units): 
\begin{eqnarray}
{\partial {\rho }}/\partial t &=&L\rho =-i[H,\rho ]+\mathcal{L}\rho ,  \notag
\\
\mathcal{L}\rho &=&\frac{1}{2}\sum_{i=1}^{n}(2\Gamma _{i}\rho \Gamma
_{i}^{\dag }-\rho \Gamma _{i}^{\dag }\Gamma _{i}-\Gamma _{i}^{\dag }\Gamma
_{i}\rho ),  \label{master}
\end{eqnarray}%
where the dissipator $\mathcal{L}$ describes the incoherent processes,
arising from system-bath interaction. We include spontaneous emission from
level $|3\rangle $ at rates $\gamma _{13}$ and $\gamma _{23}$ via Lindblad
operators%
\begin{equation}
\;\Gamma _{1}=\gamma _{13}|1\rangle \langle 3|\;,\Gamma _{2}=\gamma
_{23}|2\rangle \langle 3|.
\end{equation}%
We also include collisional relaxation between levels $|1\rangle $ and $%
|2\rangle $ at rates $\gamma _{12}$ and $\gamma _{21}$ via Lindblad
operators 
\begin{equation}
\;\Gamma _{3}=\gamma _{12}|1\rangle \langle 2|\;,\Gamma _{4}=\gamma
_{21}|2\rangle \langle 1|.
\end{equation}

\subsection{Review of open systems geometric phase}

\label{review}

To see how a geometric phase can be associated with the master equation
evolution, we follow Ref. \cite{SarandyLidar:06} and write the master
equation as 
\begin{equation}
{\partial {\rho }}/\partial t=L[\vec{R}(t)]\rho (t),  \label{eq:t-Lind1}
\end{equation}%
where $L$ depends on time only through a set of parameters $\vec{R}(t)\equiv 
\vec{R}$. These parameters will undergo adiabatic cyclic evolution in our
problem.

In the superoperator formalism, the density matrix for a quantum state in a $%
D$-dimensional Hilbert space is represented by a $D^{2}$-dimensional
\textquotedblleft coherence vector\textquotedblright\ $|\rho \rangle \rangle
=\left( \rho _{1},\rho _{2},\cdots ,\rho _{D^{2}}\right) ^{t}$ (where
$t$ denotes the transpose) and the 
Lindblad superoperator $L$ becomes a $D^{2}\times D^{2}$-dimensional
supermatrix \cite{Alicki:87}, so that the master equation (\ref{eq:t-Lind1})
can be written as linear vector equation in $D^{2}$-dimensional
Hilbert-Schmidt space, in the form $\partial |\rho \rangle \rangle /\partial
t=L[\vec{R}(t)]|\rho \rangle \rangle $. Such a representation can be
generated, e.g., by introducing a basis of Hermitian, trace-orthogonal, and
traceless operators [e.g., the $D$-dimensional irreducible representation of
the generators of su($D$)], whence the $\rho _{i}$ are the expansion
coefficients of $\rho $ in this basis \cite{Alicki:87}, with $\rho _{1}$ the
coefficient of $I$ (the identity matrix).

The master equation generates a non-unitary evolution since $L$ is
non-Hermitian. In fact, $L$ need not even be a normal operator ($L^{\dag
}L\neq LL^{\dag }$). Therefore $L$ is generally not diagonalizable, i.e., it
does not possess a complete set of linearly independent eigenvectors.
Equivalently, it cannot be put into diagonal form via a similarity
transformation. However, one can always apply a similarity transformation $S$
to $L$ which puts it into the (block-diagonal) Jordan canonical form~\cite%
{Horn:book}, namely, $L_{\mathrm{J}}=S^{-1}LS$. The Jordan form $L_{\mathrm{J%
}}$ of a $D^{2}\times D^{2}$ matrix $L$ is a direct sum of blocks of the
form $L_{\mathrm{J}}=\oplus _{\alpha =1}^{m}J_{\alpha }$ ($\alpha $
enumerates Jordan blocks), where $m\leq D^{2}$ is the number of linearly
independent eigenvectors of $L$, $\sum_{\alpha =1}^{m}n_{\alpha }=D^{2}$
where $n_{\alpha }\equiv \dim J_{\alpha }$ is the dimension of the $\alpha $%
th Jordan block, and $J_{\alpha }=\lambda _{\alpha }I_{n_{\alpha }}+K_{a}$
where $\lambda _{\alpha }$ is the $\alpha $th (generally complex-valued)
Lindblad-Jordan (LJ) eigenvalue of $L$ (obtained as roots of the
characteristic polynomial), $I_{n_{\alpha }}$ is the $n_{\alpha }\times
n_{\alpha }$ dimensional identity matrix, and $K_{a}$ is a nilpotent matrix
with elements $(K_{a})_{ij}=\delta _{i,j-1}$ ($1$'s above the main
diagonal), where $\delta $ is the Kronecker symbol. Since the sets of left
and right eigenvectors of $L$ are incomplete (they do not span the vector
space), they must be completed to form a basis. Instantaneous right $\{|%
\mathcal{D}_{\beta }^{(j)}[\vec{R}(t)]\rangle \rangle \}$ and left $%
\{\langle \langle \mathcal{E}_{\alpha }^{(i)}[\vec{R}(t)]|\}$ bi-orthonormal
bases in Hilbert-Schmidt space can always be systematically constructed by
adding $n_{\alpha }-1$ new orthonormal vectors to the $\alpha $th left or
right eigenvector, such that they obey the orthonormality condition $\langle
\langle \mathcal{E}_{\alpha }^{(i)}|\mathcal{D}_{\beta }^{(j)}\rangle
\rangle =\delta _{\alpha \beta }\delta ^{ij}$~\cite{SarandyLidar:04}. Here
superscripts enumerate basis states inside a given Jordan block ($i,j\in
\{0,...,n_{\alpha }-1\}$). When $L$ is diagonalizable, $\{|\mathcal{D}%
_{\beta }^{(j)}[\vec{R}(t)]\rangle \rangle \}$ and $\{\langle \langle 
\mathcal{E}_{\alpha }^{(i)}[\vec{R}(t)]|\}$ are simply the bases of right
and left eigenvectors of $L$, respectively. If $L$ is not diagonalizable,
these right and left bases can be constructed by suitably completing the set
of right and left eigenvectors of $L$ (which can be identified with columns
of $S$ and $S^{T}$, respectively, associated with distinct eigenvalues $%
\lambda _{\alpha }$). Then for all times $t$ 
\begin{eqnarray*}
L|\mathcal{D}_{\alpha }^{(j)}\rangle \rangle &=&|\mathcal{D}_{\alpha
}^{(j-1)}\rangle \rangle +\lambda _{\alpha }|\mathcal{D}_{\alpha
}^{(j)}\rangle \rangle , \\
\langle \langle \mathcal{E}_{\alpha }^{(i)}|L &=&\langle \langle \mathcal{E}%
_{\alpha }^{(i+1)}|+\lambda _{\alpha }\langle \langle \mathcal{E}_{\alpha
}^{(i)}|,
\end{eqnarray*}%
so that the $\{|\mathcal{D}_{\alpha }^{(j)}\rangle \rangle \}$ and $%
\{\langle \langle \mathcal{E}_{\alpha }^{(i)}|\}$ preserve the Jordan block
structure (see Appendix~A of Ref. \cite{SarandyLidar:06} for a detailed
discussion of these issues).

In order to define geometric phases in open systems, the coherence vector is
expanded in the instantaneous right vector basis $\{|{\mathcal{D}_{\beta
}^{(j)}[\vec{R}(t)]\rangle \rangle }\}$ as 
\begin{equation}
|\rho (t)\rangle \rangle =\sum_{\beta =1}^{m}\sum_{j=0}^{n_{\beta
}-1}p_{\beta }^{(j)}(t)\,e^{\int_{0}^{t}\lambda _{\beta }(t^{\prime
})dt^{\prime }}\,|\mathcal{D}_{\beta }^{(j)}[\vec{R}(t)]\rangle \rangle ,
\label{rho_supop}
\end{equation}%
where the dynamical phase $\exp [\int_{0}^{t}\lambda _{\beta }(t^{\prime
})dt^{\prime }]$ is explicitly factored out. The coefficients $\{p_{\beta
}^{(j)}(t)\}$ play the role of \textquotedblleft
geometric\textquotedblright\ (non-dynamical) amplitudes. We assume that the
open system is in the adiabatic regime, i.e., \emph{Jordan blocks associated
to distinct eigenvalues evolve in a decoupled manner} \cite{SarandyLidar:04}%
. Then: 
\begin{equation}
{\dot{p}}_{\alpha }^{(i)}\,=\,p_{\alpha }^{(i+1)}-\sum_{\beta \,|\,\lambda
_{\beta }=\lambda _{\alpha }}\sum_{j=0}^{n_{\beta }-1}p_{\beta
}^{(j)}\langle \langle \mathcal{E}_{\alpha }^{(i)}|{\dot{\mathcal{D}}}%
_{\beta }^{(j)}\rangle \rangle .  \label{pgen}
\end{equation}%
Note that, due to the restriction $\lambda _{\beta }=\lambda _{\alpha }$,
the dynamical phase has disappeared.

A condition on the total evolution time, which allows for the neglect of
coupling between Jordan blocks used in deriving Eq.~(\ref{pgen}), was given
in Ref.~\cite{SarandyLidar:04}. This condition generalizes the standard
closed-system adiabaticity condition \cite{Messiah:vol2}, from which Eq.~(%
\ref{eq:adcond}) is derived. Nevertheless, we have used the simpler
condition (\ref{eq:adcond}) in our simulations below, as it is rather
accurate in the present open system case.

For closed systems, Abelian geometric phases are associated with
non-degenerate levels of the Hamiltonian, while non-Abelian phases appear in
the case of degeneracy. In the latter case, a subspace of the Hilbert space
acquires a geometric phase which is given by a matrix rather than a scalar.
For open systems, one-dimensional Jordan blocks are associated with Abelian
geometric phases in the absence of degeneracy, or with non-Abelian geometric
phases in case of degeneracy. Multi-dimensional Jordan blocks are always
tied to a non-Abelian phase \cite{SarandyLidar:06}.

\subsubsection{The Abelian case: generalized Berry phase}

Consider the simple case of a non-degenerate one-dimensional Jordan block (a
block that that is a $1\times 1$ submatrix containing an eigenvalue of $L$).
In this case, the absence of degeneracy implies in Eq.~(\ref{pgen}) that $%
\lambda _{\beta }=\lambda _{\alpha }\Rightarrow \alpha =\beta $
(non-degenerate blocks). Moreover, since the blocks are assumed to be
one-dimensional we have $n_{\alpha }=1$, which allows for removal of the
upper indices in Eq.~(\ref{pgen}), resulting in ${\dot{p}}_{\alpha
}=-p_{\alpha }\langle \langle \mathcal{E}_{\alpha }|{\dot{\mathcal{D}}}%
_{\alpha }\rangle \rangle $. The solution of this equation is $p_{\alpha
}(t)=p_{\alpha }(0)\exp {[i\beta _{\alpha }(t)]}$, with ${\beta }_{\alpha
}(t)=i\int_{0}^{t}\langle \langle \mathcal{E}_{\alpha }(t^{\prime })|{\dot{%
\mathcal{D}}}_{\alpha }(t^{\prime })\rangle \rangle dt^{\prime }$. For a
cyclic evolution in parameter space along a closed curve $C$, one then
obtains the Abelian geometric phase associated with the Jordan block $\alpha 
$ \cite{SarandyLidar:06}: 
\begin{equation}
{\beta }_{\alpha }(C)=i\oint_{C}\langle \langle \mathcal{E}_{\alpha }(\vec{R}%
)|\vec{\bigtriangledown}|{\mathcal{D}}_{\alpha }(\vec{R})\rangle \rangle
\cdot d\vec{R}.  \label{gp_abelian}
\end{equation}%
This expression for the geometric phase bears clear similarity to the
original Berry formula, Eq.~(\ref{eq:Berry}). Note that in general ${\beta }%
_{\alpha }(C)$ can be complex, since $\langle \langle \mathcal{E}_{\alpha }|$
and $|{\mathcal{D}}_{\alpha }\rangle \rangle $ are not related by transpose
conjugation. Thus, the geometric phase may have real and imaginary
contributions, the latter affecting the visibility of the phase. As shown in
Ref. \cite{SarandyLidar:06}, the expression above for ${\beta }_{\alpha }(C)$
satisfies a number of desirable properties: it is \emph{geometric }(i.e.,
depends only on the path traversed in parameter space), it is \emph{gauge
invariant} (i.e., one cannot modify the geometric phase by redefining $%
\langle \langle \mathcal{E}_{\alpha }|$ or $|{\mathcal{D}}_{\alpha }\rangle
\rangle $ via multiplication of one of them by a complex factor); it has the
proper \emph{closed system limit} (if the interaction with the bath
vanishes, ${\beta }_{\alpha }(C)$ reduces to the usual difference of
geometric phases acquired by the density operator in the closed case).

\subsubsection{The non-Abelian case}

Ref.~\cite{SarandyLidar:06} also derived the non-Abelian open systems
geometric phase, for the case of degenerate one-dimensional Jordan blocks. A
non-Abelian geometric phase in fact arises in our STIRAP\ model when the
spontaneous emission rates are equal. However, we shall not treat this case
in the present paper.

\subsection{Solution of the STIRAP model}

Returning to the STIRAP model, let us represent the density matrix $\rho $
in terms of the coherence vector $\vec{v}$ as 
\begin{equation}
\rho =\frac{1}{N}\left[ \mathbf{1}+\sqrt{\frac{N(N-1)}{2}}\sum_{\alpha
}v_{\alpha }\Omega _{\alpha }\right] ,
\end{equation}%
where the $\Omega _{\alpha }$ are the Gell-Mann matrices \cite{Gellmann:book}%
. Writing the Lindblad equation $\dot{\rho}=L\rho $ in the $\{\Omega
_{\alpha }\}$ basis, we obtain $\dot{\vec{v}}=L\vec{v}$, where $\vec{v}=%
\frac{1}{3}[1,\sqrt{3}v_{1},\cdots ,\sqrt{3}v_{8}]^{t}$ is a nine-component
coherence vector. In the same basis we can
express the Liouville operator $L$ in the following form: 
\begin{widetext}
\begin{equation}
L=\left(\begin{array}{ccccccccc}
0&0&0&0&0&0&0&0&0\\0&-\gamma'_+&0&0&0&g_2&0&g_1&0\\0&0&-\gamma'_+&0&-g_2&0&g_1&0&0\\\frac{\gamma_-}{2}+\gamma'_-&0&0&-2\gamma'_+&0&g_1&0&-g_2&-\frac{\gamma_--\gamma'_-}{\sqrt{3}}\\
0&0&g_2&0&-\gamma_+-\frac{\gamma_{21}^2}{2}&0&0&0&0\\0&-g_2&0&-g_1&0&-\gamma_+-\frac{\gamma_{12}^2}{2}&0&0&-\sqrt{3}g_1\\0&0&-g_1&0&0&0&-\gamma_+-\frac{\gamma_{21}^2}{2}&0&0\\
0&-g_1&0&g_2&0&0&0&-\gamma_+-\frac{\gamma_{12}^2}{2}&-\sqrt{3}g_2\\\sqrt{3}\gamma_+&0&0&0&0&\sqrt{3}g_1&0&\sqrt{3}g_2&-2\gamma_+
\end{array}\right)\;,
\end{equation}
\end{widetext}
where we have used the Hamiltonian (\ref{hamil}). Here 
\begin{equation}
\gamma _{+}=(\gamma _{13}^{2}+\gamma _{23}^{2})/2,\quad \gamma _{-}=(\gamma
_{13}^{2}-\gamma _{23}^{2}),\quad \gamma _{\pm }^{\prime }=(\gamma
_{12}^{2}\pm \gamma _{21}^{2}),
\end{equation}%
and $g_{1,2}$ are given in Eq.~(\ref{pulse}).

The eigenvalues and left and right eigenvectors of $L$ can be found in terms
of the parameters $\gamma _{\pm },\gamma _{\pm }^{\prime }$ and $g_{1,2}$,
but the expressions are very complicated. The analytic determination of the
corresponding left and right eigenvectors is cumbersome, so instead we have
used a numerical procedure, which is based on the discussion presented in
subsection \ref{review} above \cite{program}.

To exhibit some of the analytic structure, we temporarily make the further
simplification that the spontaneous emission rates are equal: $\gamma
_{13}=\gamma _{23}\equiv \gamma $. This is the case, e.g., for $D_{2}$
transitions in $^{23}$Na \cite{sodium}. In addition we assume temporarily
that the collisional relaxation rates vanish:$\ \gamma _{12}=\gamma _{21}=0$%
. With these simplifications $L$ has the following three sets of eigenvalues
(the ordering of subscripts is explained below): 
\begin{eqnarray}
&&\{\lambda _{4},\lambda _{5},\lambda _{6}\}=\{0,-\gamma ^{2},-\gamma ^{2}-%
\frac{Q}{3P}+P\}\;,  \notag  \label{eq:evals} \\
&&\lambda _{1}=(-\gamma ^{2}+\frac{Q}{6P}-\frac{P}{2})+\frac{i\sqrt{3}}{2}(%
\frac{Q}{3P}+P);  \notag \\
&&\lambda _{9}=(-\gamma ^{2}+\frac{Q}{6P}-\frac{P}{2})-\frac{i\sqrt{3}}{2}(%
\frac{Q}{3P}+P)  \notag \\
&&\lambda _{2}=\lambda _{3}=\frac{1}{2}(-\gamma ^{2}+i\sqrt{Q});\text{ }%
\lambda _{7}=\lambda _{8}=\frac{1}{2}(-\gamma ^{2}-i\sqrt{Q})  \notag \\
&&
\end{eqnarray}%
where 
\begin{eqnarray}
P &=&\left( x+\sqrt{x^{2}+(Q/3)^{3}}\right) ^{1/3}\;,  \notag \\
Q &=&4(g_{1}^{2}+g_{2}^{2})-\gamma ^{4}\;,  \notag \\
x &=&\gamma ^{2}(g_{1}^{2}+g_{2}^{2})
\end{eqnarray}%
and the last set of four eigenvalues appears in two degenerate pairs.
Because of this, the corresponding open systems geometric phase is
non-Abelian (recall the discussion above), but we do not consider this case
here.

In the closed system limit ($\gamma \rightarrow 0$) $L$ becomes $-i[H,\cdot
] $ and its eigenvalues are $\epsilon _{nm}=i(E_{n}-E_{m})$ ($n,m\in 0,+,-$%
), where $E_{n,m}$ are the eigenvalues of the control Hamiltonian $H$ as
given in Eq.~(\ref{eq:E}). The grouping in Eq.~(\ref{eq:evals}) represents
this limit in the following sense: 
\begin{eqnarray}
&&\lambda _{4},\lambda _{5},\lambda _{6}\rightarrow \epsilon _{nn}=0;  \notag
\\
&&\lambda _{1}\rightarrow \epsilon _{+-}=2i\sqrt{g_{1}^{2}+g_{2}^{2}}; 
\notag \\
&&\lambda _{9}\rightarrow \epsilon _{-+}=-2i\sqrt{g_{1}^{2}+g_{2}^{2}}; 
\notag \\
&&(\lambda _{2}\rightarrow \epsilon _{+0})=(\lambda _{3}\rightarrow \epsilon
_{0-})=i\sqrt{g_{1}^{2}+g_{2}^{2}};  \notag \\
&&(\lambda _{7}\rightarrow \epsilon _{0+})=(\lambda _{8}\rightarrow \epsilon
_{-0})=-i\sqrt{g_{1}^{2}+g_{2}^{2}}.  \label{eq:limits}
\end{eqnarray}%
The corresponding eigenvectors of $L$ reduce to $|n\rangle \langle m|$. The
subscripts of the $\lambda _{\alpha }$ represents the ordering of the
eigenvalues in the closed system limit. We find that the degeneracy leading
to a non-Abelian geometric open system phase appears only when $\gamma
_{13}=\gamma _{23}$ \emph{and} $\gamma _{12}=\gamma _{21}=0$, or in the
closed system limit.

By a coordinate transformation from the control fields $g_{1,2}$ to the
angle $\theta =\arctan (g_{1}/g_{2})$ we have, similarly to the closed
system case, from the generalized geometric phase formula Eq.~(\ref%
{gp_abelian}): 
\begin{equation}
\beta _{\alpha }=i\int_{0}^{\pi /2}d\theta \langle \langle \mathcal{E}%
_{\alpha }|\frac{d}{d\theta }|\mathcal{D}_{\alpha }\rangle \rangle \;.
\label{eq:gp-open}
\end{equation}%
This expression for the phase associated with the $\alpha $th eigenvector of 
$L$ yields, in the closed system limit, not the absolute phase of each of
the adiabatic eigenstates of the system Hamiltonian, but rather their phase 
\emph{differences}. This is natural as only a phase difference is an
experimentally measurable quantity.

\section{Results and Discussion}

\label{results}

\begin{figure}[tbp]
\scalebox{0.6}{\includegraphics{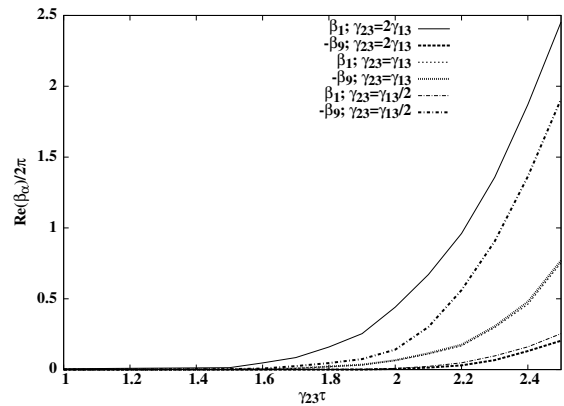}}
\caption{Spontaneous emission without collisional relaxation: Variation of
Re($\protect\beta _{1}$) (thin lines) and $-$Re$(\protect\beta _{9})$ (thick
lines), in units of $2\protect\pi $, with respect to $\protect\gamma _{23}%
\protect\tau $, for three different ratios between $\protect\gamma _{23}$
and $\protect\gamma _{13}$. The other parameters used are$:$ $\protect\gamma %
_{12}=\protect\gamma _{21}=0$, $t_{0}=4\protect\tau /3$, $g_{01}\protect\tau %
=g_{02}\protect\tau =15$, and $\protect\tau =1$.}
\label{fig1}
\end{figure}

\begin{figure}[tbp]
\scalebox{0.6}{\includegraphics{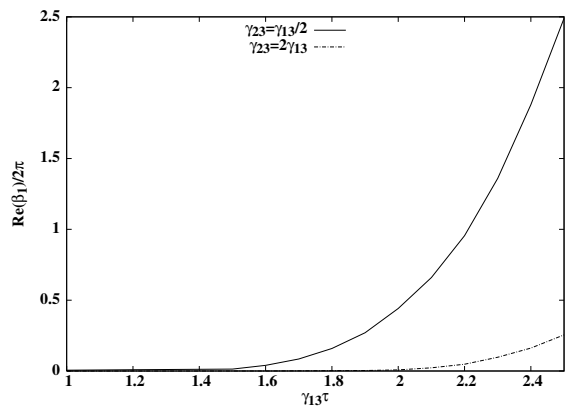}}
\caption{Effect of reversing the order of the control pulses. Now $t_0=-4%
\protect\tau/3$ and all other parameters are the same as in Fig.~\protect\ref%
{fig1}.}
\label{fig-reverse}
\end{figure}

\begin{figure}[tbp]
\scalebox{0.6}{\includegraphics{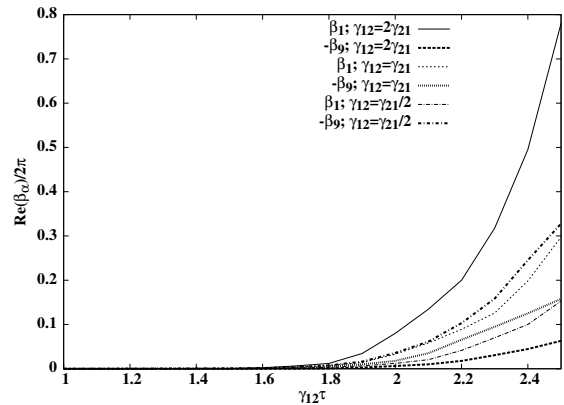}}
\caption{Collisional relaxation without spontaneous emission:\ Variation of
Re($\protect\beta _{1}$) (thin lines) and -Re$(\protect\beta _{9})$ (thick
lines) in units of $2\protect\pi $ with respect to $\protect\gamma _{12}%
\protect\tau $ for different ratios between $\protect\gamma _{12}$ and $%
\protect\gamma _{21}$. The other parameters used are: $\protect\gamma _{13}=%
\protect\gamma _{23}=0$, $t_{0}=4\protect\tau /3$, $g_{01}\protect\tau %
=g_{02}\protect\tau =15$, and $\protect\tau =1$.}
\label{fig2}
\end{figure}

\begin{figure}[tbp]
\scalebox{0.6}{\includegraphics{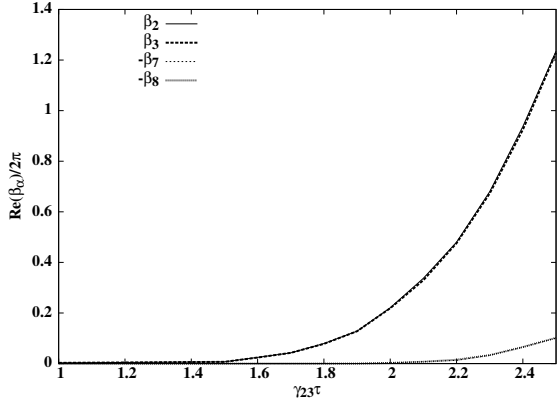}}
\caption{Spontaneous emission without collisional relaxation: Variation of
real part of phases in units of $2\protect\pi $ with respect to $\protect%
\gamma _{23}\protect\tau $ for $\protect\gamma _{23}=2\protect\gamma _{13}$.
Other parameters: $t_{0}=4\protect\tau /3$, $g_{01}\protect\tau =g_{02}%
\protect\tau =15$, and $\protect\tau =1$.}
\label{fig3}
\end{figure}

\begin{figure}[tbp]
\scalebox{0.6}{\includegraphics{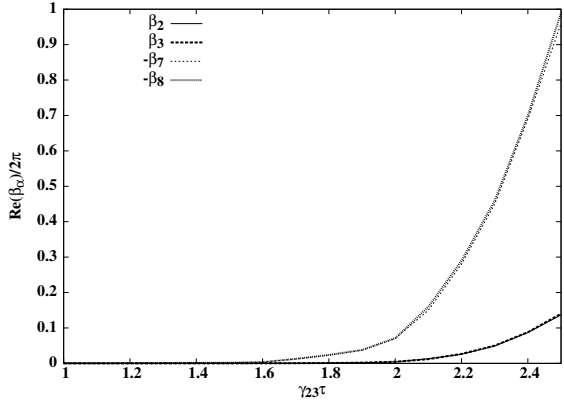}}
\caption{Spontaneous emission without collisional relaxation. Parameters the
same as in Fig.~\protect\ref{fig3}, except that $\protect\gamma _{23}=%
\protect\gamma _{13}/2$.}
\label{fig4}
\end{figure}

\begin{figure}[tbp]
\scalebox{0.6}{\includegraphics{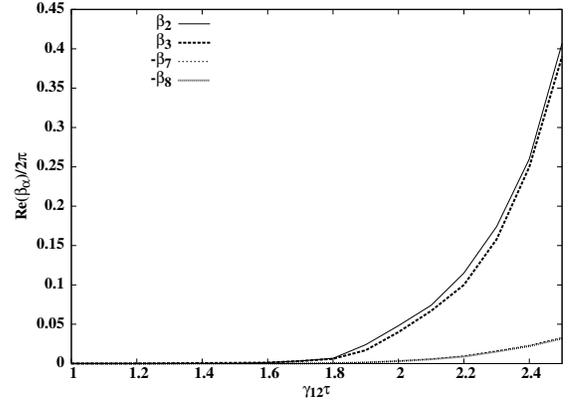}}
\caption{Collisional relaxation without spontaneous emission. Variation of
real part of phases in units of $2\protect\pi $ with respect to $\protect%
\gamma _{12}\protect\tau $ for $\protect\gamma _{23}=\protect\gamma _{13}=0$
and $\protect\gamma _{12}=2\protect\gamma _{21}$. Other parameters as in
Fig.~\protect\ref{fig3}}
\label{fig5}
\end{figure}

\begin{figure}[tbp]
\scalebox{0.6}{\includegraphics{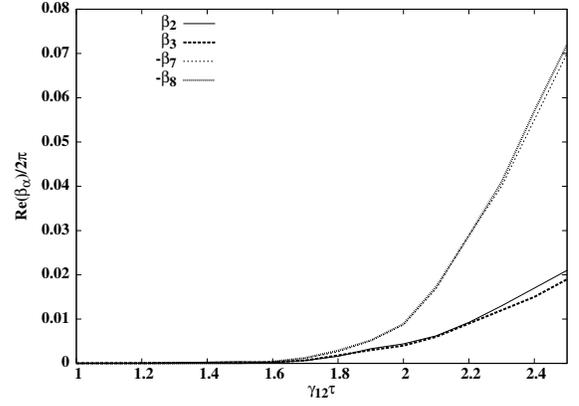}}
\caption{Collisional relaxation without spontaneous emission. Parameters the
same as in Fig.~\protect\ref{fig3}, except that $\protect\gamma _{12}=%
\protect\gamma _{21}/2$.}
\label{fig6}
\end{figure}

\begin{figure}[tbp]
\scalebox{0.6}{\includegraphics{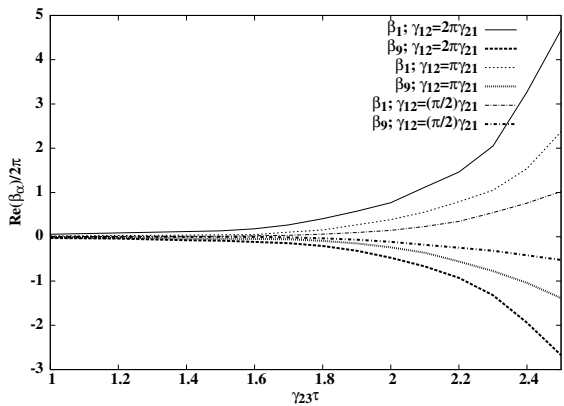}}
\caption{Spontaneous emission along with collisional relaxation. Variation
of the phases Re$(\protect\beta _{1})$ (thin lines) and Re$(\protect\beta %
_{9})$ in units of $2\protect\pi $ with respect to $\protect\gamma _{23}%
\protect\tau $ for $\protect\gamma _{23}=e\protect\gamma _{13}$ and three
different ratios between $\protect\gamma _{12}$ and $\protect\gamma _{21}$.
The other parameters used are: $t_{0}=4\protect\tau /3$, $g_{01}\protect\tau %
=g_{02}\protect\tau =15$, and $\protect\tau =1$.}
\label{fig7}
\end{figure}

We plot the real part of the open system Abelian geometric phase, i.e., Eq.~(%
\ref{eq:gp-open}), for various combinations of the spontaneous emission and
collisional relaxation rates in Figs. \ref{fig1}-\ref{fig7}. \emph{The main
finding is that the answer to the question we posed in the introduction,
\textquotedblleft Is it possible for the environment to induce a geometric
phase where there is none if the system is treated as
closed?\textquotedblright , is affirmative}. Indeed, a glance at Figs. \ref%
{fig1}-\ref{fig7} reveals that the geometric phase is non-zero, and in fact
increases with the decay rates. Moreover, we find that the imaginary part of
the geometric phase is always zero to within our numerical accuracy,
implying that the visibility of the geometric phase is unaffected in the
present case by the interaction with the environment.

In Fig.~\ref{fig1} we show the real part of the open system geometric phases 
$\beta _{1}$ and $-\beta _{9}$, for the case when collisional relaxation
vanishes ($\gamma _{12}=\gamma _{21}=0$) and there is only spontaneous
emission ($\gamma _{13},\gamma _{23}\neq 0$). Clearly, the phases increase
monotonically with the emission rates. It is interesting to note that (to
within our numerical accuracy) $\mathrm{Re}\beta _{1}=-\mathrm{Re}\beta _{9}$
when $\gamma _{13}=\gamma _{23}$. Recalling that $\lambda _{1}\rightarrow
i(E_{+}-E_{-})$ and $\lambda _{9}\rightarrow i(E_{-}-E_{+})$ [Eq.~(\ref%
{eq:limits})], this symmetry can be traced back to the difference between
the adiabatic eigenstates $|+\rangle $ and $|-\rangle $, which differ only
in the sign of the coefficient in front of the excited state $|3\rangle $
[recall Eq.~(\ref{adia})\ and that $\phi =\pi /4$]. When the spontaneous
emission rates are equal this difference in sign between the ($|3\rangle $
component of the) states $|+\rangle $ and $|-\rangle $ generates only a
difference in sign between the corresponding geometric phases, but not in
magnitude, i.e., $\beta _{1}=-\beta _{9}$.

We also note that, in spite of the symmetry between the states $|1\rangle $
and $|2\rangle $ in our model, there is an asymmetry between the curves $%
\gamma _{23}=2\gamma _{13}$ and $\gamma _{23}=\frac{1}{2}\gamma _{13}$ in
Fig.~\ref{fig1} for a given geometric phase, e.g., $\beta _{1}$. Indeed, one
might have expected a symmetry under interchange of the indices $1$ and $2$,
in the sense that, e.g., the points $\beta _{1}(\gamma _{23}=2)$ and $\beta
_{1}(\gamma _{23}=1)$ on the curves $\gamma _{23}=2\gamma _{13}$ and $\gamma
_{23}=\frac{1}{2}\gamma _{13}$ respectively, should have overlapped. That
this is not the case is because the order of the pulses $g_{2}$ (first) and $%
g_{1}$ (second) breaks the symmetry between states $|1\rangle $ and $%
|2\rangle $. Indeed, Fig.~\ref{fig-reverse} shows the results for $\beta
_{1} $ when the pulse order is reversed (now $g_{1}$ precedes $g_{2}$), and
as a consequence the order of the curves $\gamma _{23}=2\gamma _{13}$ and $%
\gamma _{23}=\frac{1}{2}\gamma _{13}$ is now reversed as well. In other
words, swapping the pulse order is equivalent to swapping the spontaneous
emission rates $\gamma _{23}$ and $\gamma _{13}$.

In Fig.~\ref{fig2} we show the real part of the open system geometric phases 
$\beta _{1}$ and $-\beta _{9}$, for the case when spontaneous emission
vanishes ($\gamma _{13}=\gamma _{23}=0$) and there is only collisional
relaxation ($\gamma _{12},\gamma _{21}\neq 0$). The results are
qualitatively similar to those in Fig.~\ref{fig1}, with the exception
that now $\mathrm{Re}\beta _{1}\neq -\mathrm{Re}\beta _{9}$ when
$\gamma 
_{12}=\gamma _{21} $. This symmetry breaking can be attributed to the fact
that the collisional relaxation operators directly connect the states $%
|1\rangle $ and $|2\rangle $, whereas these states are only connected to
second order under spontaneous emission and under the control
Hamiltonian (\ref{hamil}). The other interesting difference between
Figs.~\ref{fig1} and \ref{fig2} is that spontaneous emission only
leads to larger values of the geometric phase than collisional
relaxation only.

In Figs.~\ref{fig3}-\ref{fig6} we show the real part of the open system
geometric phases $\beta _{2},\beta_3$ and $-\beta _{7},\beta_8$. All four
phases involve the dark state $|0\rangle $ in the closed system limit.
Recall from Eq.~(\ref{eq:limits}) that the pair of eigenvalues $\lambda
_{2,3}$ becomes degenerate in the closed system limit, as does the pair $%
\lambda _{7,8}$. Figures \ref{fig3} and \ref{fig4} show the case of
vanishing collisional relaxation but non-vanishing spontaneous emission.
Figures \ref{fig5} and \ref{fig6} shows the opposite case of non-vanishing
collisional relaxation but vanishing spontaneous emission. It is interesting
to observe that when there is only spontaneous emission, as in Figs.~\ref%
{fig3} and \ref{fig4}, the phases corresponding to degenerate eigenvalues in
the closed system limit are identical to within our numerical precision. We
do not have an intuitive explanation for this symmetry, which is absent when
there is only collisional relaxation, as in Figs.~\ref{fig5} and \ref{fig6}.
On the other hand, the asymmetry between Figs.~\ref{fig3} and \ref{fig4} and
between Figs.~\ref{fig5} and \ref{fig6}, can again be attributed to the
symmetry breaking between levels $|1\rangle $ and $|2\rangle $, due to the
time ordering of the control pulses.

In Fig.~\ref{fig7} we revisit $\beta _{1}$ and $\beta _{9}$, and turn on
both spontaneous emission and collisional relaxation, and consider
irrational ratios of the various decay rates (in order to eliminate
potential accidental degeneracies due to rational ratios). Indeed, all six
curves are clearly separated, and judging by comparison to Fig.~\ref{fig1},
the effect of including both decoherence mechanisms is to increase the
magnitude of the geometric phases (i.e., the decoherence mechanisms
cooperate rather than interfere). 

\begin{figure}[tbp]
\scalebox{0.3}{\includegraphics{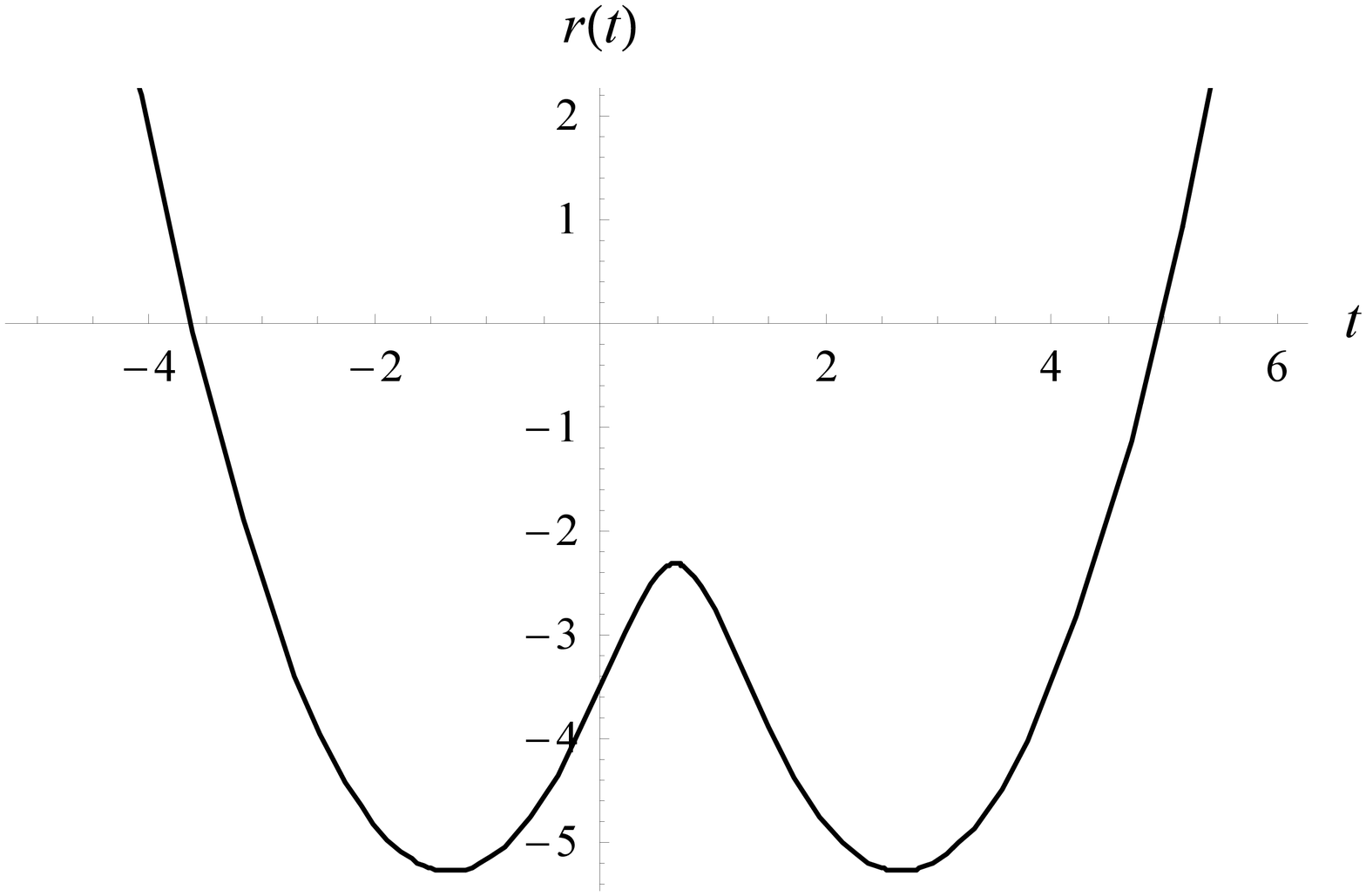}}
\caption{Natural log of the left-hand side of the adiabaticity
  condition~(\protect\ref{eq:adcond}) (i.e.,
  $r(t)=\protect\ln \protect\left( \left\vert \frac{\partial \protect\theta }{\protect\partial
  t}\protect\right \protect\vert / \protect\sqrt{g_{1}(t)^{2}+g_{2}(t)^{2}}\protect\right)$) for the parameters used in our
  simulations, i.e., $t_{0}=4\protect\tau /3$, $g_{01}\protect\tau =g_{02}%
\protect\tau =15$, and $\protect\tau =1$.}
\label{fig:adcond}
\end{figure}

\begin{figure}[tbp]
\scalebox{0.6}{\includegraphics{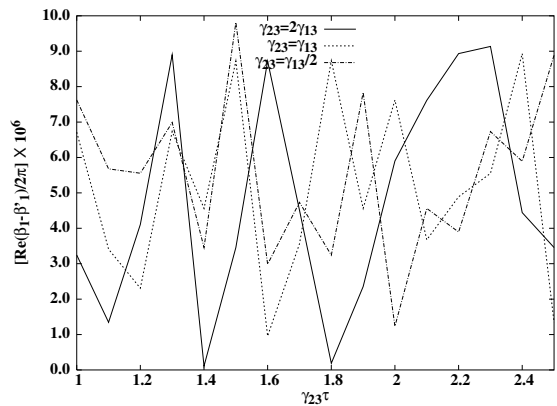}}
\caption{The phase $\protect\beta'_1$ is computed by integrating from
  $\protect\theta_1$ (corresponding to $t_1=-3.06\protect\tau$) to
  $\protect\theta_2$ (corresponding to $t_2=4.39\protect\tau$). In this
  time range the left-hand side of Eq.~(\protect\ref{eq:adcond}) $< 1/e$. Plotted is the
  variation of the difference between $\beta_1$ and $\beta'_1$ as a
  function of $\gamma_{23}\tau$ for three different ratios between $\gamma_{23}$ and $\gamma_{13}$. The other parameters are the same as in Fig.~\protect\ref{fig1}.}
\label{fig:diff}
\end{figure}

As a final note, we should point out that varying the Rabi frequencies $g_{1}
$ and $g_{2}$ along a closed cycle in parameter space (Fig.~\ref{fig:g1g2}),
i.e., letting the time $t$ vary from $-\infty $ to $+\infty $ (which we
implement in practice by integrating from $\theta =0$ to $\pi /2$), is
incompatible with Eq.~(\ref{eq:adcond}) for all times $t$ due to the
finiteness of the parameters involved. Indeed, Fig.~\ref{fig:adcond} shows
the left-hand side of Eq.~(\ref{eq:adcond}) for the parameters we have used
in our simulations. It is clear that the adiabaticity condition is satisfied
only for $-t_{1}\lesssim t\lesssim t_{2}$, where $t_{1}\approx -3$ and $%
t_{2}\approx 4.4$. However, when we repeat our calculations of the open system
geometric phase with $\theta $ varying between angles $\theta _{1}$ and $%
\theta _{2}$ corresponding to the times $t_{1}$ and $t_{2}$ (i.e., not along
a complete cycle in the $g_{1},g_{2}$ parameter space), we find -- as can be
seen in Fig.~\ref{fig:diff} -- that the effect on the geometric phase is
entirely negligible. This confirms that our choice of parameters satisfies
the adiabatic limit for all practical purposes.

\section{Conclusions}

\label{conc}

Our study of STIRAP in an open three level quantum system reveals that the
interaction with the environment can endow a system with a geometric phase,
where none existed without the interaction with the environment.
Mathematically, the vanishing geometric phase in the closed system case is
attributable to the vanishing integrand in the Berry formula. In a certain
sense this is easily understood as the result of having a geometric phase
determined by only a single parameter ($\theta $), whence no solid angle is
traced out in parameter space. It would then be natural to conclude that, by
including the interaction with the environment a non-zero solid angle is
created, implying that in the presence of decoherence motion along an
orthogonal direction in parameter space must have taken place. However, one
must be careful in accepting this explanation, since in fact the polar
angles $\theta $ and $\phi $ do not properly describe the parameter space in
our problem:\ indeed, $\theta $ varies from $0$ to $\pi /2$ (while $\phi $
is constant) and thus does not describe a closed path, while the correct
parameter space is that defined by the pulse amplitudes $g_{1}$ and $g_{2}$
(see Fig.~\ref{fig:g1g2}). Thus a proper explanation of the intriguing
effect of an environmentally induced geometric phase is still lacking and
will be undertaken in a future publication. Here we conjecture that this is
due to the non-commutativity of the driving Hamiltonian and the decohering
processes we have considered. It should be possible to test this by using
the quantum trajectories approach to the open systems geometric phase \cite%
{Fuentes-Guridi:05}. Another interesting open question is to what extent the
finding presented here can be made useful in the context of holonomic
quantum computing \cite{Zanardi:99b}, i.e., whether can one constructively
exploit the environmentally induced geometric phase for the generation of
quantum logic gates.


\begin{thebibliography}{99}
\bibitem{program} Our Matlab code for the Jordan form of an arbitrary square
matrix is available upon request.
\end{thebibliography}

\end{document}